\documentclass[aps,twocolumn,showpacs,nofootinbib,superscriptaddress]{revtex4}
\usepackage{graphicx}
\usepackage{amsmath}
\usepackage{amsfonts}
\usepackage{amssymb}
\usepackage{hyperref}

\newcommand{\ket}[1]{|{#1}\rangle}
\newcommand{\bra}[1]{\langle{#1}|}

\newcommand{\beq}{\begin{equation}}
\newcommand{\eeq}{\end{equation}}
\newcommand{\beqa}{\begin{eqnarray}}
\newcommand{\eeqa}{\end{eqnarray}}

\begin{document}

\title{Decoherence induced by a dynamic spin environment (II):\\Disentanglement by local system-environment interactions}

\author{Cecilia Cormick}
\affiliation{Departamento de F\'\i sica, FCEyN, UBA,
Ciudad Universitaria Pabell\'on 1, 1428 Buenos Aires, Argentina}

\author{Juan Pablo Paz}
\affiliation{Departamento de F\'\i sica, FCEyN, UBA,
Ciudad Universitaria Pabell\'on 1, 1428 Buenos Aires, Argentina}

\date{\today}

\begin{abstract}
This article studies the decoherence induced on a system of two qubits by local  
interactions with a spin chain with nontrivial internal dynamics (governed by an XY Hamiltonian). Special attention is payed to the transition between two limits: one in which both qubits interact with the same site of the chain and another one where they interact with distant sites. The two cases exhibit different behaviours in the weak and strong coupling regimes: when the coupling is weak it is found that decoherence tends to decrease with distance, while for strong coupling the result is the opposite. Also, in the weak coupling case, the long distance limit is rapidly reached, while for strong coupling there is clear evidence of an expected effect: environment-induced interactions between the qubits of the system. A consequence of this is the appearance of quasiperiodic events that can be interpreted as ``sudden deaths'' and ``sudden revivals'' of the entanglement between the qubits, with a time scale related to the distance between them. 
\end{abstract}

\date{\today}
\pacs{03.65.Yz, 03.67.Lx, 03.67.Mn} 
\maketitle

\section{Introduction}

Understanding the process of decoherence \cite{Zeh-1973, PazZ00, Zurek03} is not only important from a fundamental point of view but also essential to design good error correction strategies to prevent the collapse of quantum computers 
\cite{NielsenC00}.  When a composite system (for example, a quantum computer with many qubits) interacts with an environment, the process of decoherence generates disentanglement between the components of the system. Different qubits of a quantum computer would interact with environments that may sometimes be effectively independent, but in other cases the correlations in the environment may not be negligible. In such a context, it is important to understand the properties of the decoherence process and the signs in the decay of quantum coherence of an environment which is not only dynamic (i.e., it has a nontrivial evolution of its own) but also spatially  correlated. In this paper we make a step in this direction. 

In a previous paper we analyzed the decoherence induced on a single qubit by the interaction with an environment formed by a spin chain with an XY Hamiltonian \cite{CormickP-2007}. The way in which quantum coherence decays in time was found to depend strongly on the nature of the environment's Hamiltonian (see \cite{CormickP-2007} for a list of references of previous studies on spin baths). Here, we will consider a system formed by two qubits interacting with an environment modelled in the same way. Each qubit in our system will be put in local contact with separate sites of the chain. This situation is physically more realistic and interesting than the more familiar one of central qubits with homogeneous couplings to all the chain. We note, however, that our system is equivalent to a one-qubit system with non-local couplings. The decoherence process will induce a decay of the entanglement between the qubits that will depend on the distance between them. We will study different scenarios ranging from the case where both qubits interact with the same site, to the long distance situation which is similar to the action of two independent environments \cite{rossini-2007-75}. We will specially consider two limiting cases of very weak and very strong system-environment couplings, and see that the decoherence effects differ substantially. In the strong coupling regime, the chain provides a medium through which the qubits can interact, so that the entanglement between them exhibits quasiperiodic events of ``sudden deaths'' and ``sudden revivals'' \cite{yu-2004-93} with a time scale that depends on the distance between qubits. 

The structure of the paper is as follows: in Section \ref{sec:themodel} we introduce the model, defining the Hamiltonians for the system, the environment, and the coupling between them. We also present the main formulas we will use to determine the decay of quantum coherence. In Sections \ref{sec:weakcoupling} and \ref{sec:strongcoupling} we study the decay of the entanglement between the qubits as a consequence of their non-homogeneous interactions with the environment, in the cases of weak and strong couplings with the chain. Finally, in Section \ref{sec:conclusions} we summarize our results.

\section{The model}
\label{sec:themodel}

\subsection{The system, the environment, and their evolution}

We will study the decoherence induced on a system of two spin $1/2$ particles
(which we shall call ``the system'' or ``the qubits'') by the coupling to an environment formed by a chain of $N$ spin $1/2$ particles. We neglect the self-Hamiltonian of the system, and take the Hamiltonian of the environment chain as:
\beq
H_C=-\sum_j\left\{ \frac{1+\gamma}{2} X_j X_{j+1} + \frac{1-\gamma}{2} Y_j Y_{j+1} + \lambda Z_j\right\}
\eeq
where $X_j,Y_j,Z_j$ denote the three Pauli operators acting on the $j$-th site of the chain, and periodic boundary conditions are imposed. The parameter $\gamma$ determines the anisotropy in the $x-y$ plane and $\lambda$ gives a magnetic field in $z$ direction ($\gamma=1$ corresponds to the Ising chain with transverse field). 
This model is critical for $\gamma=0, |\lambda|<1$ and for $\lambda=\pm1$, which corresponds (in the limit $N\to\infty$) to a quantum phase transition from a ferromagnetic to a paramagnetic phase. Throughout the paper, we shall discuss the effects of this phase transition on the decoherence of the system.

Each of the two qubits is made to interact locally with a spin of the chain: the first qubit  ($S_1$) with the spin $1$ (this choice is arbitrary because of the periodic boundary conditions), and the second qubit ($S_2$) with the spin $n$. The interaction Hamiltonian is chosen as:
\beq
H_{int}=- g~ \left(\ket{1}\bra{1}_{S_1}Z_1 + \ket{1}\bra{1}_{S_2} Z_n\right)
\eeq 
with $\ket{0}$ and $\ket{1}$ the two eigenstates of the $Z$ Pauli operator.
Thus, if the system is in state $\ket{ab}$ ($a,b\in\{0,1\}$), the environment evolves with an effective Hamiltonian $H_{ab}$ given by:
\beq
H_{ab}=H_C - g (a Z_1 + b Z_n).
\eeq
In this way, the effect of the coupling is a change in the effective external field of spins $1$ and $n$, conditioned on the state of the system.

We assume the initial state of the universe formed by the system and the environment to be pure and separable (i.e., the environment is not correlated with the system):
\beq
\rho_{SE}(0)= \ket{\psi}\bra{\psi} \otimes \ket{E_0}\bra{E_0}
\eeq
Our goal is to study the evolution of the reduced density matrix of the two-qubit system (obtained from the state of the universe by tracing out the environment). Because of the special form of the Hamiltonian, the temporal dependence of the reduced density matrix $\rho$ can be formally obtained as follows. In the computational basis of the two--qubit system (formed by the eigenstates of the $Z$ Pauli operator of each qubit), $\rho$ can be written as:
\beq
\rho(t)=\sum_{abcd=0,1}\rho_{ab,cd}(t) \ket{ab}\bra{cd}
\eeq
The evolution of the elements of $\rho$ is given by:
\beq
\rho_{ab,cd}(t)=\rho_{ab,cd}(0) \bra{E_0}e^{iH_{cd}t} e^{-iH_{ab}t}
\ket{E_0}
\eeq
(we are using units such that $\hbar=1$). As the total Hamiltonian commutes with $Z_{S_1}$ and $Z_{S_2}$, the diagonal terms $\rho_{ab,ab}$ remain constant. 
On the other hand, each off-diagonal term in the reduced density matrix is modified by a factor smaller than 1. This factor is associated with the overlap between two states of the environment that correspond to the two different evolutions of the spin chain according to the different system states. To analyze the decoherence induced by the coupling to the spin chain we will consider the square modulus of this factor, which (after \cite{rossini-2007-75}) we shall call the Loschmidt echo: 
\beq \label{eq:echodef}
L_{ab,cd} (t) = |\bra{E_0}e^{iH_{cd}t} e^{-iH_{ab}t}\ket{E_0}|^2.
\eeq
This echo is simplified if we assume the environment to be initially in the ground state of one of the effective Hamiltonians involved. In such a case, one of the evolution operators in the expression acts trivially, and the echo is then equal to the survival probability of the initial state after being evolved with the other operator.

To find the decay of coherence with time for this problem we note that the Hamiltonians $H_{ab}$ of the chain can be mapped onto a fermion system by means of the Jordan-Wigner transformation \cite{LiebSM-1961}:
\beqa
X_j&=&exp\left\{i\pi\sum_{k=1}^{j-1}c_k^\dagger c_k\right\} (c_j+c_j^\dagger)\\
Y_j&=&i~exp\left\{i\pi\sum_{k=1}^{j-1}c_k^\dagger c_k\right\} (c_j-c_j^\dagger)\\
Z_j&=&2c_j^\dagger c_j-1. \label{eq:transfZ}
\eeqa
Using this, up to a correction term associated to boundary effects, the Hamiltonians can be written as:
\beqa
H_{ab}&=&-\sum_j (c_j^\dagger c_{j+1} +c_{j+1}^\dagger c_j) + \gamma (c_j^\dagger c_{j+1}^\dagger +c_{j+1} c_j) +\nonumber\\
&&+ \lambda_j^{(ab)} (2c_j^\dagger c_j-1) 
\eeqa
with $\lambda_j^{(ab)}=\lambda+ g(a\delta_{j,1}+b\delta_{j,n})$ (the extension to the case where the 
qubits interact with more than one site is trivial). 

The Hamiltonians $H_{ab}$ depend quadratically on the annihilation and creation operators. Therefore they can be diagonalized by linear (Bogoliubov) transformations defining new creation and annihilation operators which we will denote as $\eta^{(ab)}, \eta^{\dagger(ab)}$. Furthermore, as all these transformations are linear, the operators corresponding to different values of the labels $(ab)$ can also be connected by Bogoliubov transformations.

The factors modulating the off-diagonal terms of the reduced density matrix can be written in terms of these transformations. If the initial state of the environment $\ket{E_0}$ is the ground state of the unperturbed chain Hamiltonian $H_{00}$ we can show (see Appendix 1) that:
\beq \label{eq:mainformula}
L_{ab,00} (t) = \left|\bra{E_0}e^{-iH_{ab}t}\ket{E_0}\right|^2 = 
\left|\det(\rm{g}+\rm{h} e^{i\Lambda t})\right|^2
\eeq
Here $\Lambda$ is a diagonal $N\times N$ matrix containing the energies of the normal modes of the Hamiltonian $H_{ab}$. $\rm{g}$ and $\rm{h}$ correspond to the transformation connecting the particles that diagonalize the unperturbed Hamiltonian $H_{00}$ and the effective Hamiltonian $H_{ab}$: 
\beq \label{eq:lineartransf}
\eta_j^{(00)}=\sum_k \rm{g}_{jk}\eta_k^{(ab)}+\rm{h}_{jk}\eta_k^{\dagger(ab)}
\eeq
In this way we can express the echo as the determinant of an $N\times N$ matrix, which can be efficiently computed (the number of operations required is polynomial in the size of the chain). This formula is just a different version of the one used in \cite{rossini-2007-75}, where the Loschmidt echo was written in terms of the two-point correlators of the environment chain; our formula is improved as the matrix in \cite{rossini-2007-75} was $2N\times 2N$. 

In what follows we will analyze the decoherence process for the two-qubit system.
For this purpose we would need to compute $L_{ab,cd}$ for all values of the labels $(ab)$ and $(cd)$. The formula we presented above enables us to compute this echo when either $(ab)$ or $(cd)$ are equal to $(00)$ (provided that the initial state of the environment is an eigenstate of the Hamiltonian $H_{00}$). In order to study the decay of the other density matrix elements, we need to compute the survival probability of the initial state $\ket{E_0}$ evolving first with Hamiltonian $H_{ab}$ and then evolving backwards with $H_{cd}$. So far, we do not have a simple formula like (\ref{eq:mainformula}) enabling us to efficiently evaluate the echo in the most general case (see discussion in the Appendix). Thus, we will restrict ourselves to consider the evolution of initial states for which equation (\ref{eq:mainformula}) provides the full answer. These initial states are a superposition of two computational states of the system, one of which is the state $\ket{00}$. The case where the other state is either $\ket{01}$ or $\ket{10}$ corresponds to a situation in which one of the qubits remains in state $\ket{0}$, without interacting with the environment, while the other is affected by the chain in a way that has been extensively treated in \cite{rossini-2007-75}. 

\begin{figure}[!hbt]
\begin{center}
\includegraphics[width=0.25\textwidth]{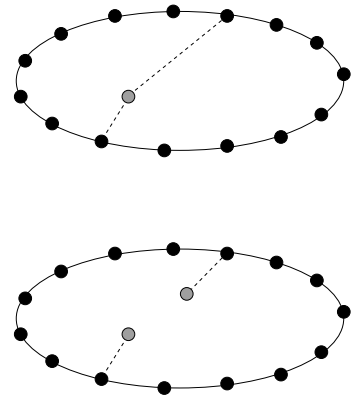}
\caption{The one-qubit system interacting with distant sites of a spin chain is equivalent to the problem of a two-qubit system prepared in the symmetric state $\alpha\ket{00}+\beta\ket{11}$, where the qubits interact locally with distant sites of the chain but not with each other.}
\label{fig:2spines}
\end{center}
\end{figure}

In order to study the effects of the environment over the entanglement in the system, we shall take the initial state to be of the form $\ket{\psi}=\alpha \ket{00}+\beta \ket{11}$. It should be noted, however, that this case is also reducible to a single-qubit system, interacting with two separate sites of the chain (which has been partially studied in \cite{rossini-2007-75}). This equivalence is illustrated in Figure \ref{fig:2spines}. But even if the two problems are equivalent, the case of separate qubits interacting locally with distant sites is a more realistic situation. It also allows us to address new interesting questions related to the entanglement decay induced by the environment, and its dependence on the distance between the sites to which the two qubits are coupled. In this way we will also generalize previous studies where the two qubits were treated as ``central qubits'' (each one interacting with equal strength with all the sites in the chain), as in \cite{jing-2006}.

\subsection{Disentanglement and decoherence}

Both the entanglement within the two-qubit system and that of the system with the chain can be analyzed by studying the Loschmidt echo (\ref{eq:mainformula}). Because of the initial state chosen, the only relevant echo we need to compute is ${L}\equiv {L}_{00,11}$. The entanglement generated between the system and the chain can be measured from the purity of the reduced density matrix of the system: while the initial state of the system is pure, as a consequence of the interaction the qubits become entangled with the chain and the state of the system becomes mixed. The purity as a function of time can be computed as:
\beq
Tr(\rho^2(t))=1-2|\alpha\beta|^2(1-L).
\eeq

The interaction with the environment will also reduces the entanglement between the two qubits. This can be most easily measured by the negativity $\mathcal N$ \cite{VidalW-2002}, obtained by partially transposing the reduced density matrix of the system, and computing the absolute sum of its negative eigenvalues (a nonvanishing negativity is an indicator of non-separability of the two-qubit state). For the initial state proposed, the negativity is: 
\beq
\mathcal N_\psi (t)= |\alpha\beta|L^{1/2}
\eeq
If the initial state contains any entanglement (e.g.
\mbox{$|\alpha\beta|\neq0$}) then the entanglement in the system decays as $L^{1/2}$, and vanishes only when the off-diagonal terms vanish. 

The relation between the decay of the off-diagonal terms and the disentanglement of the two qubits is not always so straightforward. In fact, if the initial state is a mixture of the form $\rho=(1-p)\ket{\psi}\bra{\psi}+p~\mathbb{I}/4$, the negativity is given by: 
\beq
\mathcal N(t)= \max\{0,(1-p)\mathcal N_\psi(t)-p/4\}. 
\eeq
This means that, for a given value of $p$, the state becomes disentangled when the echo is below a threshold value given by $L_{lim}(p)=[4|\alpha\beta|(-1+1/p)]^{-2}$. 
This is a well-known phenomenon that has been called ``sudden death of entanglement'' \cite{yu-2004-93}: the entanglement can vanish at finite time even if coherence decays asymptotically (or does not decay completely). This can happen here if we mix the initial state $\ket{\psi}$ with the identity, taking $p$ large enough for $L$ to reach the threshold, but small enough to have some entanglement at the beginning.

\subsection{From mutually indistinguishable to far-apart qubits}


In the following we shall analyze the behaviour of the echo for different values of the distance $d=n-1$ between the sites that interact with the qubits. More precisely, one of our interests is the study of two distinct limits and the transition between them: first, if both qubits are coupled to the environment through the same site, then they cannot be told apart by the environment. We shall call this the case of mutually indistinguishable qubits. The other limit is the one in which the qubits are coupled to the environment through very distant sites. In such a case, if the environment is sufficiently large, its effect should be equivalent to the one produced by two identical independent environments (as reported in \cite{rossini-2007-75}). Our study will enable us to see the transition between these two limits. It is worth pointing out that for the independent environment regime to exist the back-action of the qubits on the environment must be small. In the case of strong back-action, one expects the environment to induce an effective interaction between the qubits. Thus, the long distance limit would be identifiable properly in the weak coupling case, whereas the strong coupling regime would be plagued with effects associated with such environment-induced interactions (which would depend on the distance between the qubits). 

It is interesting to describe the expected behaviour for the echo in the limits mentioned above. Let us consider first the case $d=0$ where both qubits are coupled to the same site. If the coupling strength of each qubit is $ g$, then the resulting echo ${L}_{00,11}( g,t)$ is exactly the one corresponding to a single qubit interacting with one site of the chain (which we denote as ${L}_{0,1}$), with twice the interaction strength:
\beq
d = 0\ \implies \ L_{00,11}( g, t) = L_{0,1}(2 g, t).\label{dto0}
\eeq
The other limit is also easy to understand. If the two qubits are coupled to distant sites of a sufficiently large environment, we expect that each qubit will effectively be interacting with an independent environment (as mentioned above, this is true provided that the interaction between qubits mediated by the environment is indeed negligible). For independent environments it is easy to compute the echo for the two qubits (${L}_{00,11}( g, t)$) in terms of the single-qubit echo (${L}_{0,1}( g, t)$), as:
\beq
d\gg 1\ \implies \ {L}_{00,11}( g, t)\rightarrow {L}_{0,1}^2( g, t).\label{dtoinfty}
\eeq
We will investigate the decoherence in the above limits and the transition between both, analyzing how the entanglement between the qubits is affected by the fact that they are interacting with the same environment. We will show that the effective interaction induced between the qubits in the strong coupling regime manifests in the form of revivals of the echo that depend on the distance between the interaction sites.

\section{Weak coupling}
\label{sec:weakcoupling}

\begin{figure}[!hbt]
\begin{center}
\includegraphics[width=0.45\textwidth]{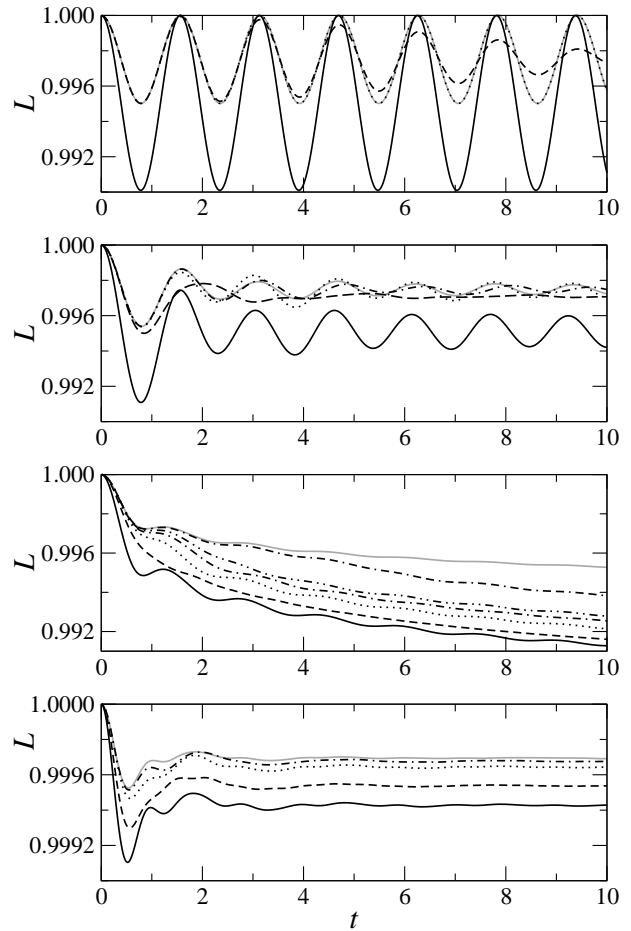}
\caption{The echo as a function of time for two qubits weakly interacting with two different sites of a chain with $N=100$, $\gamma=1$, $ g=0.1$. The plots correspond, from top to bottom, to the cases $\lambda=0$, 0.5, 1, and 1.5. The distance between sites is $d=0$ (full), 1 (dashed), 2 (dotted), and 3 (dash-dotted). In the critical case $\lambda=1$ we include also $d=4$ ($\cdot\cdot-$), and 10 ($--\cdot$); these curves are not shown in the other plots because they are intertwined with the others. For comparison we include in grey the limit of two independent environments (each of the qubits interacting with a separate chain of length $N$).}
\label{fig:peqpert_gamma1}
\end{center}
\end{figure}

Let us consider first the weak coupling case (i.e., \mbox{$ g \ll 1$}). We will study the echo for different chain Hamiltonians and distances between the perturbed sites, considering a representative case $ g=0.1$. Figure \ref{fig:peqpert_gamma1} shows the echo as a function of time for $\gamma= 1$ and for several values of $\lambda$ and $d$.

\bigskip

The case $\gamma=1, \lambda=0$ is peculiar as it can be solved analytically. The ground state for the chain corresponds simply to an eigenstate of all the Pauli operators $X_j$, and is therefore not entangled. The resulting echo is the same for all distances $d>1$. The behaviour for $\lambda\neq0$ is different depending on whether $\lambda<1$, $\lambda=1$, or $\lambda>1$. In the first case the echo oscillates with a frequency of order 1 and small amplitude; the oscillation is about a value that does not change significantly as $d$ is varied (for $d>1$). In the critical case the echo decays rapidly up to times of order unity and logarithmically afterwards, while oscillating with very small amplitude and a frequency of the same order as before. For a given fixed time, the value of the echo increases with $d$, which means the further the interaction sites, the weaker the decoherence caused by the environment. For $\lambda>1$ (but still of order $1$) the decoherence, quantified by $\mathcal D=1-L$, is an order of magnitude smaller, and the initial oscillations decay rapidly so that the echo reaches, for times of order $1$, an almost constant value. This value increases with distance, but unlike the critical case, it rapidly saturates. Thus, we observe that for $\gamma=1$, the long distance regime is reached at a relatively small number of sites: $d\ge 4$ seems to be sufficient for the non-critical cases. It is worth mentioning that we are analyzing the echo for times shorter than the ones where finite size effects become important; these effects, that appear in the form of revivals, will be considered later.

\begin{figure}[!hbt]
\begin{center}
\includegraphics[width=0.45\textwidth]{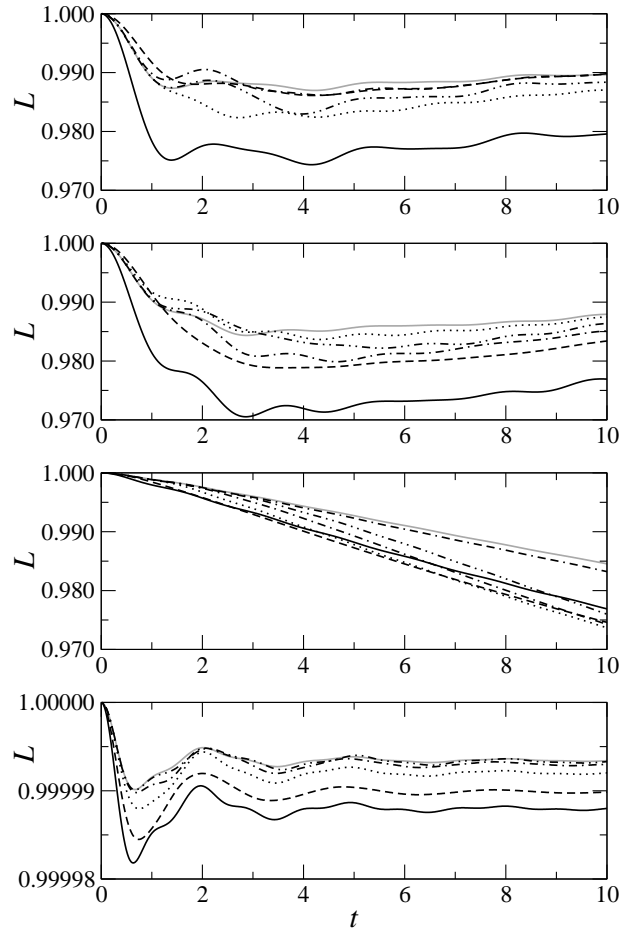}
\caption{The echo as a function of time for two qubits weakly interacting with two different sites of a chain with $N=100$, $\gamma=0.1$, $ g=0.1$. The plots correspond, from top to bottom, to the cases $\lambda=0$, 0.5, 1, and 1.5. The distance between sites is $d=0$ (full), 1 (dashed), 2 (dotted), 3 (dash-dotted), and 4 ($\cdot\cdot-$). In the critical case $\lambda=1$ we include also $d=10$ ($--\cdot$); for the sake of clarity this curve is not shown in the other plots. For comparison we include in grey the limit of two independent environments (each of the qubits interacting with a separate chain of length $N$).}
\label{fig:peqpert_gamma01}
\end{center}
\end{figure}

A similar study is presented for the case $\gamma=0.1$ in Figure \ref{fig:peqpert_gamma01}. This case is much less regular than $\gamma=1$. For $\lambda<1$ the echo oscillates with a frequency of order 1 and another one an order of magnitude smaller (but still different from $g$); the behaviour with distance is very irregular. In the critical case, the echo decays almost linearly for times up to $t\sim10$ (after which a logarithmic decay starts); at short times decoherence decreases with distance, but at longer times it is not monotonic: as $d$ is varied between 0 and $N/2$ with $t$ fixed, $\mathcal D$ increases with $d$ for small distances, and then it starts decreasing again. For $\lambda>1$, we recover a behaviour similar to the case $\gamma=1$, but $\mathcal D$ is two orders of magnitude smaller (for the other values of $\lambda$ the order of magnitude of decoherence was similar for $\gamma=1$ and $\gamma=0.1$). The long distance limit is not reached as rapidly as for $\gamma=1$. 

In almost all the non-critical cases analyzed, the interaction with the same site corresponds to the strongest decoherence (we found exceptions to this rule in cases not displayed, as $\gamma=0.1$, $\lambda=0.9$). The reason for this result, that at first sight may appear strange, is that the coupling is indeed weak. A na\"ive argument supporting this is the following: at short times one expects a Gaussian decay of the echo, of the form $\exp(-\alpha t^2)$. The time width of this decay is inversely proportional to $\sqrt{\alpha}\propto g$. For $d=0$ we have $\alpha\sim(2 g)^2$ while in the case of independent environments $\alpha\propto2 g^2$. As a consequence, more decoherence is expected in the weak coupling regime for $d=0$.

\begin{figure}[!hbt]
\begin{center}
\includegraphics[width=0.45\textwidth]{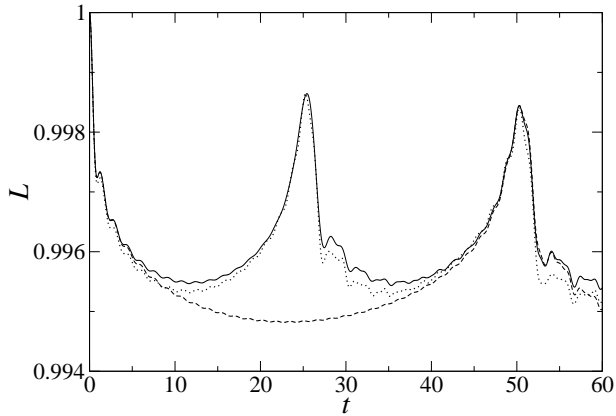}
\caption{The echo for two qubits weakly interacting with two different sites in the same chain at long times has revivals due to finite-size effects. We show the echo for $N=100$ and distance $d=N/2$ (full), and compare it with the case where each qubit interacts with one site in two separate chains of length $N$ (dashed) and $N/2$ (dotted). The parameters of the chain Hamiltonian correspond to the critical case $\lambda=1$, $\gamma=1$, and $ g=0.1$.
}
\label{fig:1vs2chains}
\end{center}
\end{figure}

The  previous study shows that environment-assisted interactions between the qubits are not strong. In the strong coupling regime, that will be studied below, such effects are much more important. However, it is interesting to see that some behaviour of this type arises in the weak-coupling scenario when studying the evolution for long times. In fact, over those longer times one expects the finite size of the environment to play a role and some recoherence to occur. The existence of revivals in the echo was indeed shown in \cite{rossini-2007-75} for the case of a single spin coupled to one site of an Ising chain. The revival time depends on the chain parameters: in the case discussed in \cite{rossini-2007-75} the revival was found at $t\simeq N/2$ for $\gamma=1$, $\lambda\gtrsim1$, and at longer times for $\lambda=0.9$. In the two-qubit case we are analyzing here, as the perturbation acts on two different sites of the chain, the revivals are related not only to the finite size of the chain but also to the finite distance between these sites. This can be seen in Figure \ref{fig:1vs2chains}: there we compare the case $d=N/2$ (interaction sites that are opposite in the chain) with the situation where the two qubits interact each with one site of two different chains.  

\vspace{35pt}

\begin{figure}[!hbt]
\begin{center}
\includegraphics[width=0.45\textwidth]{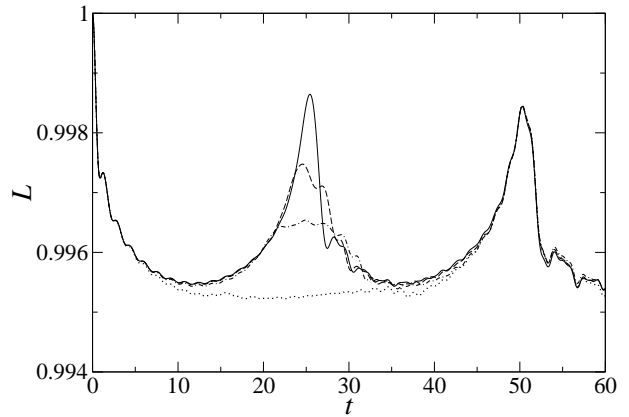}
\caption{The echo as a function of time for two qubits weakly interacting with two different sites in the same chain, with $N=100$, at distance $d=$ 50 (full), 45 (dashed), 40 (dash-dotted) and 30 (dotted). The parameters of the chain Hamiltonian correspond to the critical case $\lambda=1$, $\gamma=1$, and $ g=0.1$. The second peak is a consequence of the finite size of the chain. The first peak is associated to the distance between the qubits: it reaches its maximum for $d=N/2$, and becomes negligible for small enough $d$.}
\label{fig:longtimerevival}
\end{center}
\end{figure}

Figure \ref{fig:1vs2chains} is interesting since it shows that some of the revival peaks are clearly induced by the existence of the second interaction site (and not by the finite size of the chain itself). However, the large peak amplitude is due to the fact that we are looking at the very special case \mbox{$d=N/2$}. If we decrease the distance, the first peak becomes fainter, while the second one remains the same (Figure \ref{fig:longtimerevival}). Thus, the peak that is associated to the distance between the qubits tends to disappear when this distance is much smaller than the size of the environment (Figure \ref{fig:longtimerevival} suggests that $d<N/4$ is enough). We can then picture the qubits as evolving almost independently, even for long times. This will not be true in the regime of strong perturbations, which will be treated in the next section. 

\section{Strong coupling}
\label{sec:strongcoupling}

We consider now the strong perturbation regime, corresponding to a coupling $ g\gg1$ (we shall take $ g=50$). This limit was studied in \cite{cucchietti-2006, CormickP-2007} in connection with the problem of universality, namely, the fact that for large enough $g$ the echo displays a fast oscillation (with frequency of order $g$) with an envelope which is independent of $g$. 

\begin{figure}[!hbt]
\begin{center}
\includegraphics[width=0.45\textwidth]{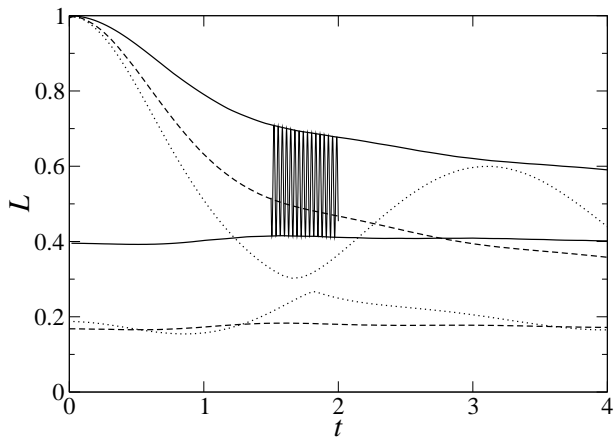}
\caption{Envelope of the echo as a function of time for two qubits strongly interacting with two different sites of a chain with $N=100$, $\lambda=0.99$, $\gamma=1$, $ g=50$. The plots correspond to the cases of distance $d=0$ (full black), 2 (dotted), and long-distance limit (dashed). For the case $d=0$ a part of the fast oscillation is included. The shape of the envelope is determined by the chain dynamics and the distance between the interaction sites. Here, in contrast to the weak coupling scenario, decoherence is stronger in the long distance limit, as the envelope for large $d$ corresponds to the square of the one for $d=0$.}
\label{fig:revivals_gamma1}
\end{center}
\end{figure}

In the regime of strong coupling we expect for qubits interacting with the same site a weaker decoherence than for two very distant qubits (behaving as if they were coupled to independent environments). This can be explained by the universal behaviour we just mentioned: in the case $d=0$ we obtain $L_{00,11}( g, t) = L_{0,1}(2 g, t)$, but this is of the same order as $L_{0,1}( g, t)$ because for large enough $g$ the envelope is independent of $g$. On the other hand, in the long distance limit $L_{00,11}( g, t) \simeq L_{0,1}^2( g, t)$, and this is smaller than $L_{0,1}( g, t)$ because the echo is smaller than 1. The comparison between $d\gg1$ and $d=0$, illustrated in Figure \ref{fig:revivals_gamma1} for $\gamma=1$, \mbox{$\lambda=0.99$}, thus leads to a result which is quite the opposite of the one obtained for weak coupling.  

There is in this regime another interesting effect, associated with the fact that the two qubits interact with the same chain. In Figure \ref{fig:revivals_gamma1} we see that, apart from the fast oscillation and the expected decay of the envelope (that are already present in the single qubit case), for $d=2$ there is a beating, or, more appropriately, an echo revival. This is a consequence of the presence of two interaction sites; indeed, the timescale of the revival is related to the distance between these two sites. This effect is not important in the weak coupling limit studied before, and it is not related to the finite size of the chain; it can instead be interpreted as an interaction between the qubits through the modes of the perturbed chain. 

The revivals can be analyzed by looking at the spectrum of the Hamiltonian contained in the matrix $\Lambda$, and the mixing of creation and annihilation operators given by the matrix $\rm{h}$, which appear in the main formula (\ref{eq:mainformula}). The unperturbed Hamiltonian has excitations with energies given by $E_k=2[1+\lambda^2+2\lambda \cos(2\pi k/N)]^{1/2}$, which lie between $2|1-\lambda|$ and $2|1+\lambda|$. There is a two-fold degeneracy associated to $\pm k$ symmetry in all but the lowest- and highest-energy excitations. These particles are delocalized in terms of the original fermion operators $c,c^\dagger$ (which are themselves non-local with respect to the spin operators, but whose population is given by the one-site magnetization (\ref{eq:transfZ})). 
When the strong effective external field is applied in two sites, two eigenvalues of order $g$ appear, which are responsible for the fast oscillation of the echo. These excitations are associated to combinations of the original fermion operators in those two sites. The remaining eigenvalues are still of the same order as before (but the degeneracy is broken). These other excitations are wave-like, but not entirely delocalized: they can be split in two groups, corresponding to excitations between the interaction sites, and outside this region. The matrix $\rm{h}$ is related to the populations of the particles that diagonalize the perturbed Hamiltonian in the ground state of the unperturbed chain. These can, using equation (\ref{eq:lineartransf}), be written as:
\beq
\bra{E_0} \eta_j^{\dagger(11)} \eta_k^{(11)} \ket{E_0} =(\rm{h}^t\rm{h})_{jj}.
\eeq 
We find that the most populated levels correspond to the lowest energy excitation (occupying the outside region), the excitations in the interaction sites, and those lying in the inside region. The ``localized'' excitations are associated to the rapid echo oscillation, and the lowest energy excitation to an oscillation with time scale growing as $N$, while the beating in the echo is given by the frequency of the lowest energy excitation in the inside region.

If we take $N\to\infty$ but keep $d$ fixed, we will find a quasiperiodic behaviour of the echo, with falls and revivals of the envelope governed by the modes of the chain contained between the qubits. As mentioned in Section \ref{sec:themodel}, for initial mixed states the echo decay can lead to sudden death of the entanglement. In the same way, now the beating associated to the interaction through the chain can also lead to a sudden revival of this entanglement.

\begin{figure}[!hbt]
\begin{center}
\includegraphics[width=0.45\textwidth]{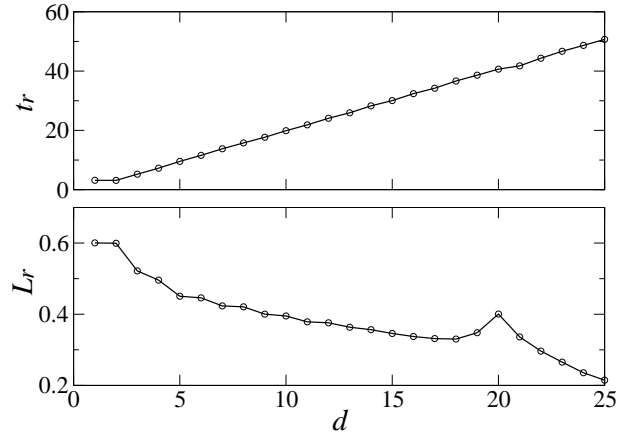}
\caption{Time of revival and value of the peak as a function of distance, for an Ising chain with $N=100$, $\lambda=0.99$, \mbox{$g=50$}. The revivals are determined by the chain dynamics and the distance between the interaction sites; the revival time is linear with $d$ with slope approximately equal to 2, while the height of the revival decreases as $d^{-1/4}$; at $d \gtrsim N/5$, finite-size effects appear inducing alterations of this behaviour.}
\label{fig:revivals_vs_d}
\end{center}
\end{figure}

Let us now consider the time of the revival, $t_r$, and the amplitude of the peak, $L_r$, as  functions of the distance $d$ between the interaction sites. We consider first the case $\lambda=0.99$, $\gamma=1$, $N=100$, which is close enough to the phase transition for $H_{00}$ and $H_{11}$ to belong to different phases. The echo for $d=0$ has no revivals until effects due to the finite size of the chain appear. For $d>1$, the revival time increases with distance, while the height of the peak  decreases. In each case, new peaks appear at multiples of $t_r$. Figure \ref{fig:revivals_vs_d} displays the dependence on distance of time and height of the first revival: the revival time is linear, which suggests a constant-velocity propagation of signals in the chain, while the value of $L_r$ decays as a power law, $L_r\propto d^{-1/4}$. The Figure exhibits also a peculiar feature associated with the parity of $d$: when $d$ is even the echo is slightly increased over the general trend, and it gets smaller for odd $d$. For distances $d \gtrsim N/5$ the effects associated with the finite size of the chain appear altering the regular trend. 

\vspace{35pt}

\begin{figure}[!hbt]
\begin{center}
\includegraphics[width=0.45\textwidth]{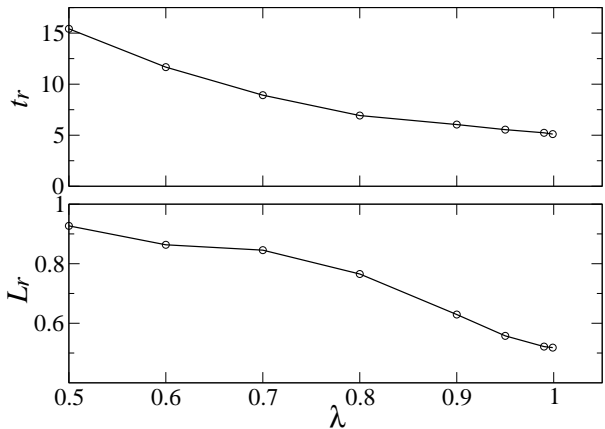}
\caption{Time of revival and value of the peak as a function of lambda, for $N=100$, $d=3$, $ g=50$. $\lambda$ is varied between 0.5 and 0.999; for $\lambda>1$ the revival disappears.}
\label{fig:revivals_vs_lambda}
\end{center}
\end{figure}

The revival peaks are not a consequence of the phase transition in the environment; they also appear for values of $\lambda<1$ such that no phase transition is involved. But for $\lambda>1$ the revival is no longer there, as a consequence of the change in the ground state properties. The times of revival and the values of the peaks are shown in Figure \ref{fig:revivals_vs_lambda} for $\lambda$ between $0.5$ and $0.999$, with $d=3$ (maintaining $\gamma=1$, $N=100$). 

\vspace{35pt}

The power-law decay of the revival peaks in the case $\lambda=0.99$ might seem to be related to the way in which the correlations decay in the spin chain. In fact, for the critical Ising chain one can show that the slowest decay, corresponding to $\langle X_iX_{i+d} \rangle$, goes as $d^{-1/4}$ \cite{BarouchMcCoy-1971, Pfeuty70}. But examining the echo far away from the phase transition (we took $\lambda=0.5$) we found a slower decay of the revival peak with $d$ (even when the correlations decay faster). Furthermore, the dependence of $t_r$ with $d$ in this case is not linear, but grows exponentially. Thus, $t_r$ seems to be related to transport properties of the chain that are modified by varying the external field. This should serve as a warning not to take too seriously the simple image of revivals as the manifestation of a spin wave propagating with constant velocity along the chain (the energies of the modes of the chain between the qubits do not generally scale as $1/d$).

\begin{figure}[!hbt]
\begin{center}
\includegraphics[width=0.45\textwidth]{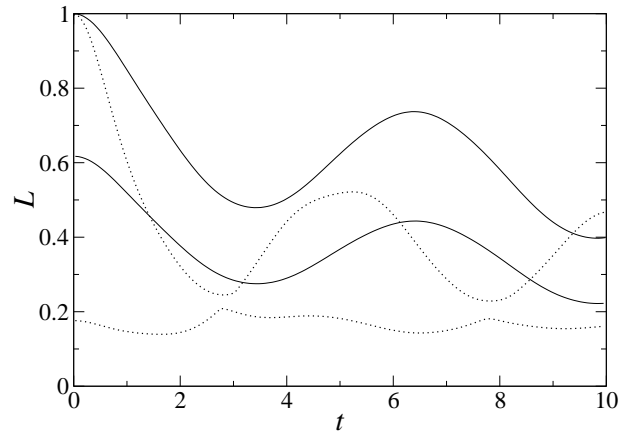}
\caption{Envelope of the echo as a function of time for two qubits strongly interacting with two different sites at distance $d=3$ of a chain with $N=100$, $\lambda=0.99$, $ g=50$. The plots correspond to the cases $\gamma=0.1$ (full) and 1 (dotted). The shape of the envelope is determined by the chain dynamics and the distance between the interaction sites.}
\label{fig:revivalsd3}
\end{center}
\end{figure}

We might also consider changes in the anisotropy parameter $\gamma$; as an example, in Figure \ref{fig:revivalsd3} we compare the envelopes for $\gamma=1$ and $\gamma=0.1$ (with $d=3$, $g=50,~\lambda=0.99$, and $N=100$). The shapes of the envelopes are different, and also the heights and times of the revival peaks differ. We studied the revival time $t_r$ and the value of $L_r$ as functions of the distance $d$ for $\lambda=0.99$ and different values of $\gamma$ between 0.1 and 1. We found that once more, the dependence of $t_r$ with $d$ is not generally linear; for $\gamma=0.1$ it becomes approximately quadratic. Besides, the height of the revival peaks decreases with $\gamma$. The dependence of $L_r$ with $d$ maintains the power-law decay but with a power that varies from approximately -$0.25$ for $\gamma=1$ to something between $-0.3$ and $-0.35$ for $\gamma=0.1$. Anyway, it should be noted that the peaks are less well-defined for large $d$ and small $\gamma$, so that these values are merely estimates. 

\vspace{35pt}

\section{Conclusions}
\label{sec:conclusions}

Here we will summarize our results. We have studied the evolution of a system of two qubits that interact locally with different sites of an XY chain acting as an environment. For this purpose we have analyzed the Loschmidt echo, which is the square modulus of the factor that determines the reduction of the off-diagonal terms in the reduced density matrix of the system. Following the ideas in \cite{rossini-2007-75} and \cite{cozzini-2006}, we have derived a formula for the evolution of the echo which can be evaluated efficiently. The entanglement within the system and that of the system with the chain were connected to the value of this echo. We found that for the initial state proposed, $\ket{\psi}=\alpha\ket{00}+\beta\ket{11}$ (with $\alpha\beta\neq0$), the entanglement in the system only vanishes when the off-diagonal terms are completely suppressed. This is not the case if we mix the initial state $\ket{\psi}$ with the identity; then the entanglement will reach zero once that the echo is below a certain value that is related to the amount of mixing and the coefficients $\alpha, \beta$.

It was our aim to study the differences in the echo as the distance between the qubits was varied, ranging from the case when both interacted with the same site to the opposite situation when they were very far apart. The cases we have analyzed show a rich variety of results. In the weak perturbation regime, we notice a very fast approach to the long distance limit, and observe that decoherence typically decreases with distance. Strong system-environment interactions, on the contrary, give rise to a different kind of behaviour, in which decoherence is larger for long distance. This strong coupling regime is characterized by fast oscillations with a short-time decay quite independent of the distance between qubits, but with a beating that provokes quasi-periodic revivals of the entanglement in the system. This effect can be interpreted as an interaction of the qubits through the chain, and its time scale is determined by the modes of the chain in the region between the qubits. The dependence of the revival time with distance is different according to the values of the parameters in the chain Hamiltonian: we have observed a linear dependence for the almost-critical case $\gamma=1, \lambda=0.99$, but also quadratic and exponential growths for other values of $\gamma$ and $\lambda$. The beatings we found correspond to $\lambda<1$; when the value of $\lambda$ was increased over the critical value (with $\gamma=1$ fixed), the observed revivals disappeared. 

We note that in the strong coupling regime the fast oscillation of the echo continually provokes decays and revivals of the entanglement between qubits. This is just a dynamical transfer of this entanglement back and forth from the system to the environment, with a frequency given by the coupling constant $g$. The true loss of entanglement is produced by the decay of the oscillation. If the initial state is mixed, this decay can lead to true sudden death of entanglement. When the distance between qubits is short, the sudden death can be followed by a sudden revival, as the chain allows for mediated interactions between the qubits. For very long distances, entanglement loss becomes irreversible, unless we consider times which are long enough for the finite size of the environment to become manifest.

\section{Appendix}

In this section we show how the formula (\ref{eq:mainformula}) for the echo can be obtained. Taking into account the relation (\ref{eq:lineartransf}) between the operators that diagonalize the Hamiltonians,
\beq
\eta_j^{(0)}=\sum_k\rm{g}_{jk}\eta_k^{(1)}+\rm{h}_{jk}\eta_k^{\dagger(1)} \nonumber
\eeq
the two different vacuum states $\ket{E_0}$ (for the unperturbed Hamiltonian) and $\ket{E_0'}$ (for the perturbed one) can be connected by \cite{chung-2001-64}:
\beq \label{eq:vacuumrelation}
\ket{E_0} \propto \exp\left\{\frac{1}{2} \vec \eta^{\dagger(1)} G \vec \eta^{\dagger(1)}\right\} \ket{E_0'} 
\eeq
with $G=-\rm{g}^{-1}\rm{h}$ (for the sake of simplicity, we shall first assume that $\rm{g}$ is invertible, and sketch the most general case afterwards). The echo can then be calculated from:
\beqa 
L (t) &=& |\bra{E_0} e^{-iH_1t} \ket{E_0}|^2 \nonumber\\ &\propto& |\bra{E_0'} e^{-\frac{1}{2} \vec \eta G \vec \eta} e^{-it\vec\eta^\dagger\Lambda\vec\eta} e^{\frac{1}{2} \vec \eta^\dagger G \vec \eta^\dagger} \ket{E_0'}|^2
\eeqa
where in the last expression $\Lambda$ is a diagonal matrix containing the energies corresponding to the different particles, and all super-indices are omitted as all states and matrices refer to the perturbed Hamiltonian $H_1$. By introducing two identities in terms of fermionic coherent states between the exponentials and integrating two times, we obtain \cite{NegeleO-1987}:
\beqa 
L (t) &\propto& \Bigg|\int d\alpha_1\ldots d\alpha_N d\beta_1^*\ldots d\beta_N^* \nonumber\\
&& \exp\left\{\frac{1}{2} \vec \alpha G \vec \alpha+\vec\beta^*e^{it\Lambda}\vec\alpha-\frac{1}{2} \vec \beta^* G \vec \beta^*\right\}\Bigg|^2
\eeqa
where $\vec\alpha, \vec\beta^*$ are Grassman $N$-tuples. This is a Gaussian integral, that can be solved to:
\beq \label{eq:Gaussianinttodet} 
L (t) \propto \left|\det
\begin{pmatrix}
G &-e^{it\Lambda}\\
e^{it\Lambda} & -G
\end{pmatrix}
\right|
\eeq
Using properties of the determinant and the facts that $\Lambda$ is diagonal and $\rm{g}\pm \rm{h}$ is orthogonal, and imposing that $L(0)=1$, the result can be rewritten in its final form (\ref{eq:mainformula}):
\beq 
L (t) = \left|\det(\rm{g}+\rm{h} e^{i\Lambda t})\right|^2 \nonumber
\eeq

In case $\rm{g}$ is not invertible, the relation (\ref{eq:vacuumrelation}) between the vacuum states can be generalized by using intermediate sets of operators in the following way: by the singular value decomposition, ${\rm g}=UDV$ with $D$ diagonal, $U,V$ orthogonal. We define new fermionic operators $\xi_j^{(0)}=\sum_kU_{kj}\eta_k^{(0)}$, $\xi_j^{(1)}=\sum_kV_{jk}\eta_k^{(1)}$. The linear transformation between the $\xi^{(i)}$, $\xi^{\dagger(i)}$ now has $D$ instead of $\rm{g}$, and the vacuum states are the same because the transformation does not mix creation and annihilation operators. We can assume $D_j=0$ for $j\leq j_1$, and the remaining eigenvalues to be nonzero. By interchanging particles with antiparticles ($\xi_j^{(0)}\leftrightarrow\xi_j^{\dagger(0)}$) for every index such that $D_j=0$ we obtain a linear transformation with an invertible matrix. The calculation of the echo follows the same steps as before, except that it is necessary to treat indices $j\leq j_1$ separately. After the Gaussian integration, we are left with an expression of the form (\ref{eq:Gaussianinttodet}) in which only indices $j>j_1$ appear. The desired result can be achieved by conveniently introducing some rows and columns in the matrix, in such a way to include the parts of the matrices with $j\leq j_1$ without changing the value of the determinant \cite{cozzini-2006}.

The present derivation cannot be easily extended to more general cases, with two different evolution operators as in (\ref{eq:echodef}), for in that case there are three sets of operators that must be related. The steps taken in the above calculation then lead to a result that involves a sum over an exponential number of terms, and so cannot be efficiently evaluated.


\end{document}